\documentclass[12pt]{article} % use larger type; default would be 10pt

\usepackage[utf8]{inputenc} % set input encoding (not needed with XeLaTeX)

\usepackage{geometry} % to change the page dimensions
\usepackage{graphicx, subfig} % support the \includegraphics command and options

\usepackage{chessboard,xskak}

\usepackage{booktabs} % for much better looking tables
\usepackage{array} % for better arrays (eg matrices) in maths
\usepackage{fancyhdr} % This should be set AFTER setting up the page geometry
\usepackage{cite}

\usepackage{amsmath, amsfonts, amsthm, amssymb, amscd}
\usepackage{indentfirst}

\usepackage[pdftex,bookmarksnumbered]{hyperref}

\newtheorem{theorem}{Theorem}

\title{A Note on the Computational Complexity of Selfmate and Reflexmate Chess Problems}
\author{Zhujun Zhang \thanks{E-mail: zhangzhujun1988@163.com} \\Government Data Management Center of \\Fengxian District, Shanghai, China}

\date{August 10, 2022} % Activate to display a given date or no date (if empty),
\begin{document}
\maketitle

\begin{abstract}

A selfmate is a Chess problem in which White, moving first, needs to force Black to checkmate within a specified number of moves. 
The reflexmate is a derivative of the selfmate in which White compels Black to checkmate with the added condition that if either player can checkmate, they must do that (when this condition applies only to Black, it is a semi-reflexmate).
We slightly modify the reduction of EXPTIME-hardness of Chess and apply the reduction to these Chess problems.
It is proved that selfmate, reflexmate, and semi-reflexmate are all EXPTIME-complete.

\end{abstract}

\section{Introduction}

In standard Chess, each player's goal is to checkmate the opponent’s king and win the game.
However, in a selfmate Chess problem, the player who checkmates loses.
So White tries to force Black to checkmate, and Black tries to avoid that in a selfmate problem.
Certainly, all moves in the game must accord with the rules of Chess \cite{chessrule}.
In this note, we discuss the computational complexity of whether White has a force win in selfmate problems.
The reflexmate is a derivative of selfmate.
In a reflexmate problem, each player compels the opponent to checkmate with the added condition that if either player can checkmate they must do that.
When this added condition applies only to Black, it is a semi-reflexmate problem.
The computational complexity of reflexmate and semi-reflexmate is also discussed in this note.

The computational complexity of Chess and some variants of Chess was studied in the last decades.
In 1981, Fraenkel and Lichtenstein \cite{chessexptimecomplete} proved that whether White has a forced win in standard Chess is EXPTIME-complete.
In 1983, Storer \cite{chesspspacehard} proved that whether White has a forced win within $k$ moves (mate-in-$k$) is PSPACE-complete where $k$ is polynomial in the board size $n$.
In 2002, Tsukiji and Yamaguchi \cite{kingchasechess} discussed the computational complexity of the king chase problem in Chess.
In 2012, Brumleve, Hamkins, and Schlicht \cite{infinitechessmateinn} proved that the mate-in-$k$ problem of Chess on an infinite board is decidable.
In 2020, Brunner et al. \cite{retrogradehelpmatechess} \cite{subwayrushhourchess} proved that retrograde and helpmate Chess problems are both PSPACE-complete.
Recently Aravind, Misra, and Mittal \cite{singleplayerchess} proved that a single-player variant of Chess, called Solo Chess, is NP-complete.

Other mathematics of Chess was also studied.
Elkies \cite{cgtchess1} \cite{cgtchess2} applied combinatorial game theory (CGT) to Chess.
The are two ways to play a two-player combinatorial game in CGT, normal play and mis\`ere play.
In normal play the player who makes the last move wins, and in mis\`ere play the player who makes the last move loses.
If we ignore draws such as stalemate, standard Chess is played in normal play, and selfmate is played in mis\`ere play.
Shitov \cite{chessnumber} \cite{chessdiameter} showed that chess number is exponential in the board size, i.e. there exist two legal positions that require exponential many moves to go between.

\section{Selfmate and Reflexmate Chess Problems}

We use two concrete examples to explain selfmate and reflexmate Chess problems.
The first example is a selfmate Chess problem from \emph{Wikipedia} \cite{selfmate}.

%图例
\begin{figure}[htbp]
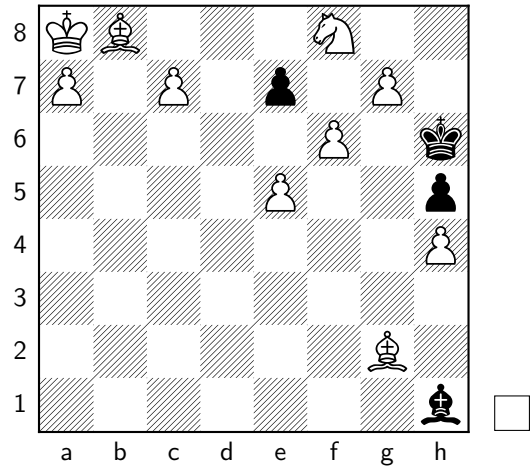
%%图
	\centering  %插入的图片居中表示
	\scalebox{1}{%
		\setchessboard{showmover=true}
		\chessboard[
		maxfield=h8,
		setpieces={
			Ka8,kh6,
			Bb8,Bg2,bh1,Nf8,
			Pa7,Pc7,Pe5,Pf6,Pg7,Ph4,pe7,ph5
		}]
	}
	\caption{A selfmate problem.}  %图片的名称
	\label{selfmateproblem}   %标签，用作引用
\end{figure}

Figure \ref{selfmateproblem} is a selfmate in two by Wolfgang Pauly in 1912.
In the problem, White to move, and she should force Black to checkmate in Black's second move.
If White can force Black to play Bxg2\#, White wins and the problem is solved.

We consider some improper plays for White:

(1) 1.Kb7 allows 1.... Bxg2+ then White fails to selfmate on move two;

(2) If White moves her bishop on square g2, the Black bishop on square h1 could move;

(3) If White moves her knight, the Black king could move;

(4) 1.g8=N\# checkmates Black so that White loses immediately;

(5) 1.g8=Q or 1.g8=R allows 1.... Bxg2+ then White queen or rook has to capture the Black bishop;

(6) 1.g8=B also allows 1.... Bxg2+ then White has to play 2.Bd5;

(7) 1.f7 or 1.fxe7 allows 1.... Kxg7;

(8) 1.e6 allows 1.... exf6 and then 2.... f5;

(9) 1.c8=Q, 1.c8=B, and 1.c8=R are not good since after 1.... Bxg2+ the White queen, bishop, or rook has to interpose with 2.Qb7, 2.Qc6, 2.Bb7, or 2.Rc6.

The only move by which White can force Black to checkmate on or before move two is 1.c8=N.
Then there are three variations: 1.... Bxg2\# is a selfmate; 1.... exf6 2.exf6 Bxg2\# is a selfmate; 1.... e6 allows 2.g8=B (the e6 pawn blocks 3.Bd5), forcing 2.... Bxg2\# and selfmate.

The second example is a reflexmate Chess problem from John Nunn's book \emph{Solving in Style} \cite{solvinginstyle}.

%图例
\begin{figure}[htbp]
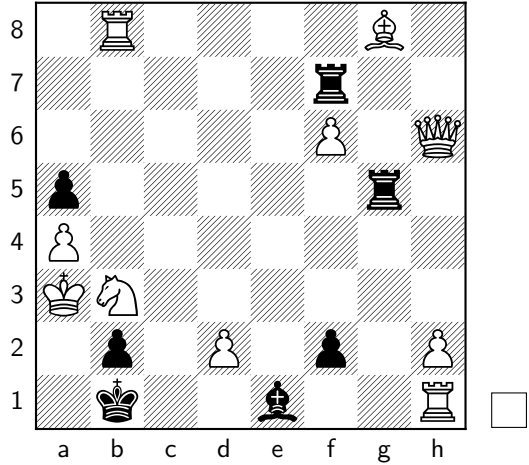
%%图
	\centering  %插入的图片居中表示
	\scalebox{1}{%
		\setchessboard{showmover=true}
		\chessboard[
		maxfield=h8,
		setpieces={
			Ka3,kb1,Qh6,
			Bg8,be1,Nb3,
			Rb8,Rh1,rf7,rg5,
			Pa4,Pd2,Pf6,Ph2,pa5,pb2,pf2
		}]
	}
	\caption{A reflexmate problem.}  %图片的名称
	\label{reflexmateproblem}   %标签，用作引用
\end{figure}

Figure \ref{reflexmateproblem} is a reflexmate in two by J. Burbach in 1978.
In the problem, White to move, and she should make that Black has one or more checkmate moves on or before Black's second move, then Black has to checkmate since the rules of reflexmate require that either player can checkmate they must do that.
Note that, Black has the threat of Rg6 now, which forces White to play Qxg6\#.
To nullify the threat and reflexmate, White should play 1.d3 intending Rb4 then Black has the checkmate move axb4\#.
There are three variations:

(1) 1.... Kc2 allows 2.Na1+ then Black has checkmate moves 2.... bxa1=Q\# and 2.... bxa1=R\#;

(2) 1.... Rb5 allows 2.Na1 then Black has checkmate moves 2.... bxa1=Q\# and 2.... bxa1=R\#, and note that White should not play 2.Nc1 since 2.... bxc1=Q+ and 2.... bxc1=B+ do not checkmate because White could play 3.Qxc1;

(3) 1.... Rb7 allows 2.Nc1 then Black has checkmate moves 2.... bxc1=Q\# and 2.... bxc1=B\#, and note that White should not play 2.Na1 since 2.... bxa1=Q+ and 2.... bxa1=R+ do not checkmate because White could play 3.Ba2.

\section{Chess is EXPTIME-complete}

Fraenkel and Lichtenstein \cite{chessexptimecomplete} proved that Chess is EXPTIME-complete.
In this section, we briefly introduce their reduction.
They reduced the game $G_3$, one of the EXPTIME-hard formula games introduced in \cite{exptimegames}, to $n \times n$ Chess.

$G_3$ is a two-player formula game where players move alternately.
A position of $G_3$ is a 4-tuple $(\tau, \text{I-LOSE} (X,Y), \text{II-LOSE} (X,Y), \alpha )$ where $\tau \in \{1,2\}$, $X = \{x_1, x_2, ...\}$, $Y = \{y_1, y_2, ...\}$, I-LOSE and II-LOSE are formulas in 12DNF, and $\alpha$ is an $(X \cup Y)$-assignment.
Player I (II) moves by changing the value assigned to exactly one variable in $X$ ($Y$) (i.e. passing is not allowed).
Player I (II) loses if the formula I-LOSE (II-LOSE) is true after some move of player I (II).
More precisely, player I can move from $(1, \text{I-LOSE} (X,Y), \text{II-LOSE} (X,Y), \alpha )$  to $(2, \text{I-LOSE} (X,Y), \text{II-LOSE} (X,Y), \alpha ')$ if and only if $\alpha '$ differs from $\alpha$ in the assignment to exactly one variable in $X$ and $\text{I-LOSE} (X,Y)$ is false under the assignment $\alpha '$.
The moves of player II are defined symmetrically.
Deciding whether player I has a forced win in $G_3$ is EXPTIME-complete.

%图例
\begin{figure}[htbp]%%图
	\centering  %插入的图片居中表示
	\includegraphics[width=0.8 \linewidth]{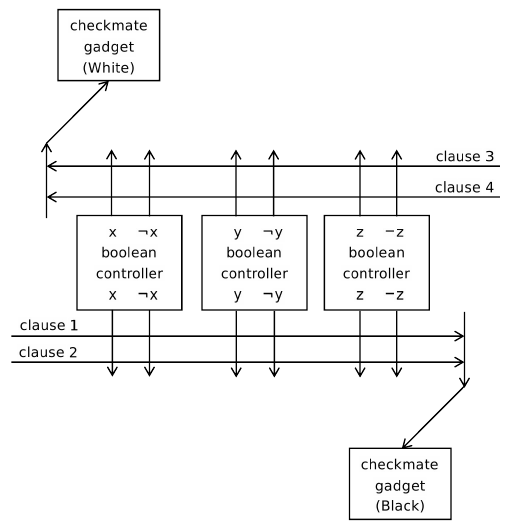}  %插入的图，包括JPG,PNG,PDF,EPS等，放在源文件目录下
	\caption{Chess is EXPTIME-complete.}  %图片的名称
	\label{chessisexp}   %标签，用作引用
\end{figure}

Fraenkel and Lichtenstein used Chess to simulate the game $G_3$, and they proved that White has a forced win in standard Chess if and only if the player I has a forced win in the game $G_3$.
Suppose that there are three variables ($x,y,$ and $z$) in the instance of game $G_3$.
Formula I-LOSE contains two clauses (clause 1 and clause 2), and formula II-LOSE also contains two clauses (clause 3 and clause 4).
Figure \ref{chessisexp} shows the structure of the reduction.
For each variable, there is a boolean controller gadget composed of the constant number of Chess pieces.
In each boolean controller, four exits are corresponding to the assignment of variables.
White (Black) moves rooks in the boolean controller gadgets, which simulates that player I (II) changes the value of variables in $G_3$.
When formula II-LOSE is true and White to move, White queens in the boolean controller gadgets could leave the boolean controller from the upper exits corresponding to the assignment of variables, and the queens can move up to the gadgets of clause 3 and clause 4.
Then one of these White queens can move up and into the checkmate gadget (White) so that the Black king is checkmated.
The condition of that formula I-LOSE is true is similar.

%图例
\begin{figure}[htbp]
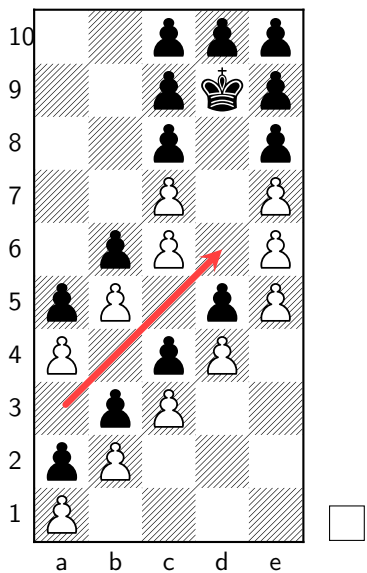
%%图
	\centering  %插入的图片居中表示
	\scalebox{1}{%
		\setchessboard{showmover=true}
		\chessboard[
		maxfield=e10,
		setpieces={
			kd9,
			Pa1,Pb2,Pc3,Pd4,Pe5,Pe6,Pe7,
			Pa4,Pb5,Pc6,Pc7,
			pa2,pb3,pc4,pd5,
			pa5,pb6,
			pc8,pc9,pc10,pd10,pe8,pe9,pe10
		},
		pgfstyle=straightmove,
		arrow=stealth,
		linewidth=.25ex,
		padding=1ex,
		color=red!75!white,
		pgfstyle=straightmove,
		shortenstart=0ex,
		showmover=true,
		markmoves={a3-d6}]
	}
	\caption{Checkmate gadget for White.}  %图片的名称
	\label{checkmategadget}   %标签，用作引用
\end{figure}

Figure \ref{checkmategadget} shows a checkmate gadget for White.
In the gadget, no pieces could move.
When a White queen enters the gadget from square a3 and moves to d6, White checkmate the Black king and wins.
The checkmate gadget for Black could be constructed similarly.
So White has a forced win in standard Chess if and only if the player I has a forced win in the game $G_3$, therefore Chess is EXPTIME-hard.
On the other hand, Chess could be solved by a simple brute-force search algorithm, so Chess is in EXPTIME.
Combining these two facts, it is proved that Chess is EXPTIME-complete.

\section{Complexity of Selfmate and Reflexmate}

Obviously, selfmate, reflexmate, and semi-reflexmate are all in EXPTIME.
So if we could prove the EXPTIME-hardness of these problems, we could prove them to be EXPTIME-complete at once.
We slightly modify the reduction of EXPTIME-hardness of Chess described in the last section to prove the hardness of selfmate, reflexmate, and semi-reflexmate.
Note that whether Player I wins or not in the game $G_3$ only depends on whether White queens enter the checkmate gadget before Black queens do.
Therefore we can design selfmate, reflexmate, and semi-reflexmate gadgets to replace the checkmate gadgets in the reduction.
Then we just need to show whether player I wins or not in the game $G_3$ only depends on whether White queens enter the selfmate, reflexmate, or semi-reflexmate gadget before Black queens do.

\subsection{Complexity of Reflexmate}

We need to construct a reflexmate gadget for White in which when a White queen enters the gadget Black has one or more checkmate moves.
The reflexmate gadget for Black could be constructed similarly.

%图例
\begin{figure}[htbp]
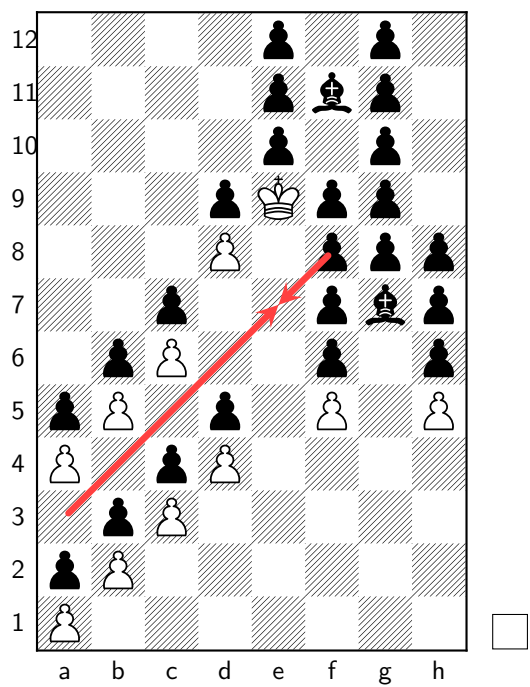
%%图
	\centering  %插入的图片居中表示
	\scalebox{1}{%
		\setchessboard{showmover=true}
		\chessboard[
		maxfield=h12,
		setpieces={
			Ke9,
			Pa1,Pb2,Pc3,Pd4,Pa4,Pb5,Pc6,Pd8,
			pa2,pb3,pc4,pd5,pa5,pb6,pc7,pd9,
			pe10,pe11,pe12,pf6,pf7,pf8,pf9,
			pg8,pg9,pg10,pg11,pg12,ph6,ph7,ph8,
			Pf5,Ph5,
			bf11,bg7
		},
		pgfstyle=straightmove,
		arrow=stealth,
		linewidth=.25ex,
		padding=1ex,
		color=red!75!white,
		pgfstyle=straightmove,
		shortenstart=0ex,
		showmover=true,
		markmoves={a3-e7,f8-e7}]
	}
	\caption{Reflexmate gadget for White.}  %图片的名称
	\label{reflexmategadget}   %标签，用作引用
\end{figure}

Figure \ref{reflexmategadget} shows a reflexmate gadget for White.
In the gadget, no pieces could move.
Suppose that a White queen enters this gadget from square a3 and moves to square e7, then the Black pawn on f8 could capture the White queen so that Black checkmates since the Black bishop on square g7 attacks the White king.

A reflexmate gadget for Black could be constructed symmetrically.
Then we could replace checkmate gadgets in the reduction of EXPTIME-hardness of Chess with reflexmate gadgets.
So White has a forced win in the reflexmate problem if and only if the player I has a forced win in the game $G_3$.
Thus we obtain the following result:

\begin{theorem}
	Reflexmate is EXPTIME-complete.
\end{theorem}

\subsection{Complexity of Selfmate}

We need to construct a selfmate gadget for White in which when a White queen enters the gadget Black has to checkmate.
So the White king must be in the White selfmate gadget.
However, how can White force Black to checkmate in this gadget?
The most common forcing moves in Chess are check moves, so White should check in the White selfmate gadget, which means that the Black king is also in this gadget.
Therefore we combine the White and Black selfmate gadgets into one gadget.

%图例
\begin{figure}[htbp]
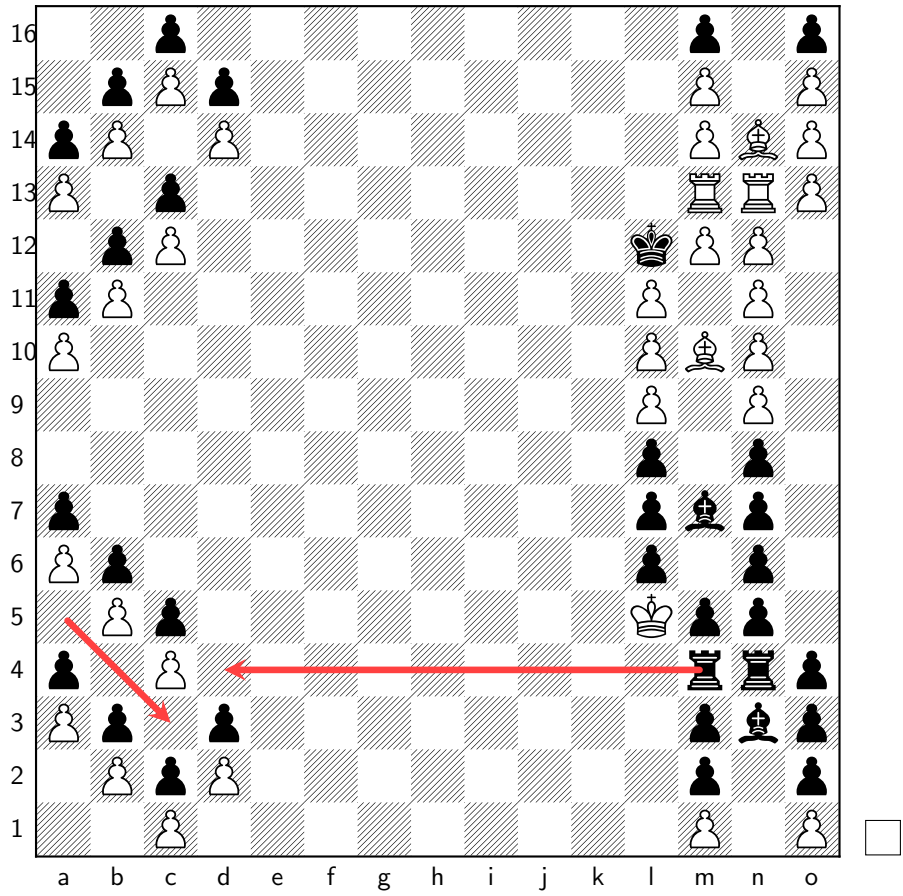
%%图
	\centering  %插入的图片居中表示
	\scalebox{1}{%
		\setchessboard{showmover=true}
		\chessboard[
		maxfield=o16,
		setpieces={
			Kl5,kl12,
			Pa3,Pb2,Pc1,Pd2,Pa6,Pb5,Pc4,
			pa4,pb3,pc2,pd3,pa7,pb6,pc5,
			Pa10,Pb11,Pc12,Pa13,Pb14,Pc15,Pd14,
			pa11,pb12,pc13,pa14,pb15,pc16,pd15,
			pl6,pl7,pl8,pn5,pn6,pn7,pn8,
			pm2,pm3,pm5,po2,po3,po4,
			Pm1,Po1,
			bm7,bn3,
			rm4,rn4,
			Pl9,Pl10,Pl11,Pn9,Pn10,Pn11,Pn12,
			Pm12,Pm14,Pm15,Po13,Po14,Po15,
			pm16,po16,
			Bm10,Bn14,
			Rm13,Rn13
		},
		pgfstyle=straightmove,
		arrow=stealth,
		linewidth=.25ex,
		padding=1ex,
		color=red!75!white,
		pgfstyle=straightmove,
		shortenstart=0ex,
		showmover=true,
		markmoves={a5-c3,m4-d4}]
	}
	\caption{Selfmate gadget.}  %图片的名称
	\label{selfmategadget}   %标签，用作引用
\end{figure}

Figure \ref{selfmategadget} shows a selfmate gadget.
In the gadget, no pieces could move except two rooks on squares m4 and m13.
But if either player moves one of these rooks, she or he checkmates and loses immediately. 
Suppose that a White queen enters this gadget from square a5 and moves to square c3, the only legal move for Black is moving the rook on square m4 to d4, so that Black checkmates since the Black bishop on square n3 attacks the White king.

Then we could replace checkmate gadgets in the reduction of EXPTIME-hardness of Chess with a selfmate gadget.
So White has a forced win in the selfmate problem if and only if the player I has a forced win in the game $G_3$.
Thus we obtain the following result:

\begin{theorem}
	Selfmate is EXPTIME-complete.
\end{theorem}

Note that, a selfmate gadget could not substitute for reflexmate gadgets in the last section.
Since no pieces could move in reflexmate gadgets, while two rooks could move in a selfmate gadget though moving these rooks leads to checkmate.
The two players both have checkmate moves in a selfmate gadget, so the player to move must checkmate and lose at once because of the rules of reflexmate, which fails to simulate the game $G_3$.

\subsection{Complexity of Semi-reflexmate}

Like the selfmate gadget, we combine the White and Black semi-reflexmate gadgets into one gadget.
We need to construct a semi-reflexmate gadget in which:

(1) When a White queen enters the gadget Black has one or more checkmate moves;

(2) When a Black queen enters the gadget White has to checkmate.

We use reflexmate and selfmate gadgets to achieve a semi-reflexmate gadget.

%图例
\begin{figure}[htbp]
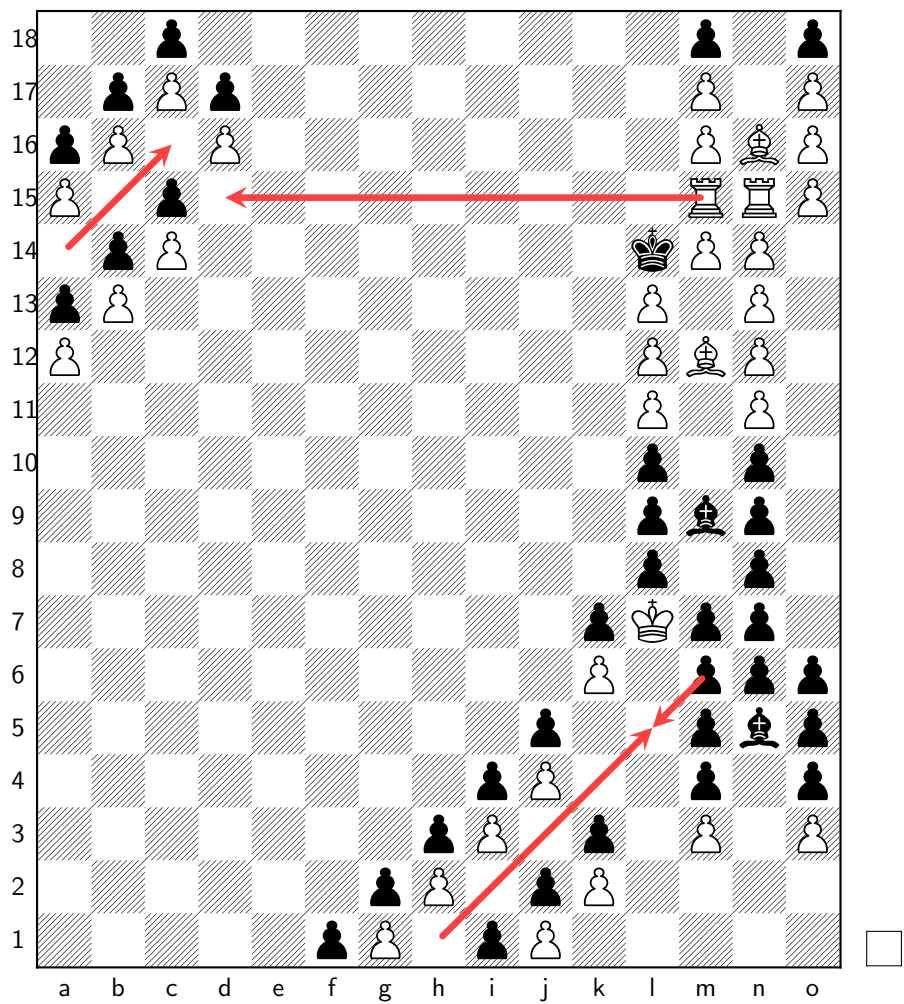
%%图
	\centering  %插入的图片居中表示
	\scalebox{1}{%
		\setchessboard{showmover=true}
		\chessboard[
		maxfield=o18,
		setpieces={
			Kl7,kl14,
			Pa12,Pb13,Pc14,Pa15,Pb16,Pc17,Pd16,
			pa13,pb14,pc15,pa16,pb17,pc18,pd17,
			pl8,pl9,pl10,pn7,pn8,pn9,pn10,
			pm4,pm5,pm7,po4,po5,po6,
			Pm3,Po3,
			bm9,bn5,
			pm6,pn6,
			Pl11,Pl12,Pl13,Pn11,Pn12,Pn13,Pn14,
			Pm14,Pm16,Pm17,Po15,Po16,Po17,
			pm18,po18,
			Bm12,Bn16,
			Rm15,Rn15,
			Pg1,Ph2,Pi3,Pj4,Pk6,Pj1,Pk2,
			pf1,pg2,ph3,pi4,pj5,pi1,pj2,pk3,pk7
		},
		pgfstyle=straightmove,
		arrow=stealth,
		linewidth=.25ex,
		padding=1ex,
		color=red!75!white,
		pgfstyle=straightmove,
		shortenstart=0ex,
		showmover=true,
		markmoves={h1-l5,m6-l5,a14-c16,m15-d15}]
	}
	\caption{Semi-reflexmate gadget.}  %图片的名称
	\label{semireflexmategadget}   %标签，用作引用
\end{figure}

Figure \ref{semireflexmategadget} shows a semi-reflexmate gadget.
In the gadget, no pieces could move except the rook on square m15.
But if White moves the rook, she checkmates and loses immediately. 
Suppose that a White queen enters this gadget from square h1 and moves to square l5, the Black pawn on m6 could capture the White queen so that Black checkmates since the Black bishop on square n5 attacks the White king.
Suppose that a Black queen enters this gadget from square a14 and moves to square c16, the only legal move for White is moving the rook on square m15 to d15, so that White checkmates since the White bishop on square n16 attacks the Black king.

Then we could replace checkmate gadgets in the reduction of EXPTIME-hardness of Chess with a semi-reflexmate gadget.
So White has a forced win in the semi-reflexmate problem if and only if the player I has a forced win in the game $G_3$.
Thus we obtain the following result:

\begin{theorem}
Semi-reflexmate is EXPTIME-complete.
\end{theorem}

%参考文献

\bibliographystyle{plain}
\bibliography{ref}

\end{document}